# Atomic-scale spin sensing of a 2D d-wave altermagnet *via* helical tunneling

Zhuying Wang[1#], Shuikang Yu[1#], Xingkai Cheng[2,3#], Xiaoyu Xiao[4#], Wanru Ma[1,5], Feixiong Quan[1], Hongxi Song[1], Kunming Zhang[1], Yunmei Zhang[1], Yitian Ma[1], Wenhao Liu[6], Priti Yadav[6], Xiangbiao Shi[4], Zhijun Wang[7], Qian Niu[1], Yang Gao[1], Bin Xiang[4*], Junwei Liu[2,3*], Zhenyu Wang[1,5*] and Xianhui Chen[1,5,8*]

[1]Department of Physics, University of Science and Technology of China, Hefei, Anhui 230026, China
[2] Department of Physics, The Hong Kong University of Science and Technology, Hong Kong, China
[3]IAS Center for Quantum Matter, The Hong Kong University of Science and Technology, Hong Kong, China
[4] Department of Materials Science & Engineering, Anhui Laboratory of Advanced Photon Science and Technology, University of Science and Technology of China, Hefei, China
[5] Hefei National Laboratory, University of Science and Technology of China, Hefei, 230088, China
[6] Department of Physics, The University of Texas at Dallas, Richardson, Texas 75080, USA
[7] Beijing National Laboratory for Condensed Matter Physics, and Institute of Physics, Chinese Academy of Sciences, Beijing 100190, China
[8] Collaborative Innovation Center of Advanced Microstructures, Nanjing 210093, China
[#]These authors contributed equally to this work
*Correspondence and requests for materials should be addressed to B. X. (binxiang@ustc.edu.cn), J. L.( liuj@ust.hk), Z.W. (zywang2@ustc.edu.cn) or X.-H.C. (chenxh@ustc.edu.cn) .

**Altermagnetism simultaneously possesses nonrelativistic spin responses and zero net magnetization, thus combining advantages of ferromagnetism and antiferromagnetism. This superiority originates from its unique dual feature, i.e., opposite-magnetic sublattices in real space and alternating spin polarization in momentum space enforced by the same crystal symmetry. Therefore, the determination of an altermagnetic order and its unique spin response inherently necessitates atomic-scale spin-resolved measurements in real and momentum spaces, an experimental milestone yet to be achieved. Here, via utilizing the helical edge (hinge) modes of a higher order topological insulator as the spin sensor, we realize spin-resolved scanning tunneling microscopy which enables us to pin down the dual-space feature of a layered *d*-wave altermagnet, $KV_2Se_2O$. In real space, atomic-registered mapping demonstrates the checkerboard antiferromagnetic order together with density-wave lattice modulation, and in momentum space, spin-resolved spectroscopic imaging provides a direct visualization of *d*-wave spin splitting of the band structure. Critically, using this new topology-guaranteed spin filter we directly reveal the unidirectional, spin-polarized quasiparticle excitations originating from the crystal symmetry-paired X and Y valleys around opposite magnetic sublattices simultaneously --the unique spin response for *d*-wave altermagnetism. Our experiments establish a solid basis for the exploration and utilization of altermagnetism in layered materials and further facilitate access to atomic-scale spin sensing and manipulating of 2D quantum materials.**

Manipulating the spin degree of freedom drives a wide field of condensed-matter research and device functionalities, ranging from spintronics (1-4) to quantum information processing (5-7). Conventionally, there are two strategies to this end: the first one relies on relativistic spin splitting due to spin–orbit coupling (SOC) as in systems lacking inversion symmetry (8, 9), and the second one utilizes nonrelativistic Zeeman splitting as in ferromagnets with broken time-reversal symmetry. Recently, a special class of antiferromagnets (AFMs) has been proposed, in which spin splitting is induced by exchange interactions and can reach the electronvolts scale even in the absence of SOC (10-28). These special spin-splitting AFMs, particularly the colinear types termed altermagnets (19,20,28-35), enable nonrelativistic spin responses with zero net magnetization, thus making them promising platforms for the next generation of information and quantum devices.

The defining feature of altermagnetism lies in the specific crystal $C$ (rotation and/or mirror), which connects opposite-spin sublattices in real space and enforces $C$-paired spin-momentum locking (SML) in momentum ($k$) space (16,19-23,28-36). Consequently, altermagnets possess symmetry-related local spin density in real space, together with alternating spin-splitting between $C$-paired $k$-points in $k$-space (28,33,37-42), as illustrated in Figure 1a-b. However, experimental studies to date have relied primarily on global measurements of the consequent phenomena (28,29,32-42), and the definitive feature of altermagnetism has evaded these spatially averaged probes due to potentially unidentified magnetic structure, the presence of multiple domains, lattice distortions, nonstoichiometry and strains (30, 34, 42-49), which has led to many contradictory conclusions (33,46-49). On the other hand, spin-polarized (SP) spectroscopic imaging with a scanning tunneling microscope (STM) remains highly challenging for quantum materials with complex surface conditions, since its spin polarization critically depends on the magnetic state of only a few atoms at the tip apex (50). Yet to date, landmark spin-resolved experiments—necessary both for achieving atomic-scale real-space resolution to clarify the local environment and opposite magnetic moments of the two sublattices, and for obtaining $k$-space resolution to reveal the corresponding spin-split band structure—have remained out of reach.

In this work, we turn topology into a metrological resource for atomic-scale spin sensing: the helical edge (hinge) modes of a higher-order (HO) topological insulator (TI) act as an intrinsic, topology-guaranteed spin filter for quantum electronic tunneling (Fig. 1c), enabling highly reliable spin-resolved STM and spectroscopic imaging for complex quantum materials. Specifically, when quantum tunneling is dominated by the helical edge modes with time-reversal-symmetry ($\mathcal{T}$)-paired SML, electrons injected into or launched from a HOTI $Bi_4Br_4$ tip carry opposite spins enforced by their helical nature. As a stringent benchmark, we apply this helical tunneling spin probe to the layered altermagnetic material $KV_2Se_2O$ and directly image the crystalline symmetry $C$ and the defining alternating spin texture in both real and momentum space, which unambiguously pin down its two-dimensional (2D) $d$-wave altermagnetic nature. Moreover, we directly detected the unique spin responses of altermagnetism at the atomic scale, i.e., spin-polarized quasiparticle excitations originating from the $C$-paired X and Y valleys around different magnetic sublattices, which establishes altermagnets as ideal platforms to study multiple degrees of freedom interactions among spin, valley, sublattice, and momentum.

**Functionalization of the tip with ideal helical edge modes**

Among the spin-sensitive techniques, SP-STM stands out as a prevalent tool capable of imaging magnetic phenomena down to the atomic level (50-52). Traditionally, this ability lies in the use of a tip with magnetic atoms/clusters at its apex, which renders the tunneling current sensitive to the relative orientation between the tip's magnetization and that of the sample (Supplementary Note 1; 50). This 'ferromagnetic' approach to functionalize a SP tip necessitates decent training and the assistance of an external magnetic field to retain the tip's magnetization. Here, we exploit a different route by utilizing topological helical edge modes as the spin sensor (Fig. 1c), which are the hallmark of 2D quantum spin Hall insulators (QSHIs) with a $\mathbb{Z}_2$ topological index (53,54) and can also emerge in 3D crystals, for example, at certain hinges in a HOTI with $\mathbb{Z}_4$ index (55, 56). The $\mathcal{T}$-paired SML of these helical edge modes mandates that electrons with opposite spins propagate in *opposite* directions along the same edge, which is highly stable due to the protection of nontrivial topology (Fig. 1d). Thus, when these propagating helical modes on one edge govern the tunneling process, electrons injected into or launched from the tip will carry opposite spins (tied to the opposite group velocities), which can be controlled by applying a negative or positive bias to the sample (Fig. 1e).

We demonstrate that this previously overlooked idea can come to fruition by choosing suitable material platforms. Bismuth bromide, $Bi_4Br_4$, is the one at point: its monolayer has been established as a 2D QSHI (57), and the bulk form, α-$Bi_4Br_4$, consisting of two layers with one of these flipped by 180° in the unit cell (AB stacking; see Fig. 2a), is suggested to be a helical HOTI with an inversion-based indicator $\mathbb{Z}_4 = 2$ (Methods; 58,59). α-$Bi_4Br_4$ is highly unique for functionalizing a SP tip in three accepts. First, it has a large bulk gap (approximately 0.28 eV) spanning the Fermi energy (Fig. 2b), enabling the helical edge modes to solely dominate the low-energy electron tunneling and to probe magnetism over a large energy window. Second, owing to the quasi 1D structure, the edge mode propagates along the chain, but its wave function is spatially confined to atomic-length scales in the other directions, resulting in weak intra- and inter-layer hybridizations. Experimentally, these edge modes can largely survive on narrow terraces with widths larger than 5 nm (Extended data Fig. 1) and are robust against vertical stacking as revealed at bilayer steps (Fig. 2c, d). Owing to the weak interlayer coupling, α-$Bi_4Br_4$ also belongs to the family of 3D QSHIs, which hold helical edge modes encircling each layer despite a tiny hybridization gap (see Methods; 60, 61). These multiple helical edge modes give rise to large electron tunneling channels. Third, the helical edge mode exhibits a single Fermi surface with linear dispersion (58, 59, 61), thus allowing for a more straightforward description of electrons' tunneling that links their spins to the group velocities.

We fabricated $Bi_4Br_4$ tips (Fig. 2e) from their bulk crystals *via* focused ion beam (FIB) based nanofabrication techniques (Methods and Extended Data Fig. 2) and characterized them on the Au (111) surface. We find that the obtained differential conductance (*dI/dV*) spectra fall into two categories: one with an insulating gap of approximately 0.25 eV and the other exhibiting metallic behavior with a suppression of the local density of states (DOS) at $E_F$ (Fig. 2f and Extended Data Fig. 3), both of which resemble the spectra of bulk α-$Bi_4Br_4$ measured with PtIr tips (Fig. 2b). Since a *dI/dV* spectrum represents a convolution of the tip and sample density of states, and metallic tips and surfaces host featureless density of states at low energies, we attribute these spectra to two types of Bi$_4$Br$_4$ tips: bulk-state tips and helical-edge-mode tips. This surprising observation suggests that the electronic structure of Bi$_4$Br$_4$ remains largely unchanged after FIB nanofabrication. Interestingly, the two statuses are switchable *in situ* by applying an electric pulse, and helical-edge-mode tips can achieve excellent atomic resolution (Extended Data Fig. 3).

For benchmarking of the spin-polarization, we use helical-edge-mode tips to image the bicollinear spin order in the antiferromagnet Fe$_{1.02}$Te (51, 62). As shown in Fig. 2g, one can immediately notice a stripe-like modulation with double periodicity (referred to bicollinear spin order) on top of the square Te lattice, which is absent in the topographies using PtIr tips (Supplementary Note 3). Strikingly, the stripe order, detected in the same field of view (FOV) from *dI/dV* mapping, shows a contrast reversal at opposite biases (Fig. 2h). This phenomenon is a direct consequence of the reversed direction of tunneling detecting the opposite-spin-polarized density of the sample (Fig. 1e), which is the hallmark for the tunneling of helical states (62).

**Crystal symmetry and density wave state in KV$_2$Se$_2$O**

Having clarified the key features of the helical-edge-mode tip, we proceed to investigate a van der Waals layered altermagnet candidate, KV$_2$Se$_2$O (Methods and Extended Data Fig. 4). It crystallizes in a tetragonal structure (*P4/mmm*) with alternating stacking of V$_2$Se$_2$O and intercalated K layers along the *c* axis (Fig. 3a, inset). The core feature is a 2D Lieb lattice consisting of two V atoms and one O atom for monolayer V$_2$Se$_2$O (63), in which the two V sublattices can form the 2D Néel AFM order (16; Supplementary Note 7). These two opposite-magnetic sublattices are thus linked by a mirror symmetry in the Se-O-Se plane (Fig. 3a) but cannot be transformed to each other by any translation operation. This symmetry allows for *d*-wave spin splitting in the electronic structure even without SOC and leads to a nonzero spin-conserved current due to anisotropic spin-polarized conductivity (15, 16). However, the unidentified phase transition at approximately 100 K (40,64-66), the debated magnetic structure (40,67) and the non-stoichiometry (63) can affect the crystal and spin symmetries and make identification elusive; thus, atomic scale insight is urgently needed.

Cleavage in KV$_2$Se$_2$O breaks the K layer and usually exposes surfaces with disordered K atoms. We find that cryo-cleaving can lead to large, ordered surfaces that are half-covered by K atoms. These surfaces are charge neutral and atomically flat but with either $\sqrt{2} \times \sqrt{2}$ or $2 \times 1$ reconstruction of the K atoms (Extended Data Fig. 5). Here, we focus on the $\sqrt{2} \times \sqrt{2}$ surface (Fig. 3a) because it does not have any external anisotropy and possesses the same symmetry as KV$_2$Se$_2$O. The *dI/dV* spectra show pronounced spatial variations: in the defect-free regions, the spectrum exhibits a hard gap with a size of ~20 meV; this gap becomes narrower somewhere and eventually collapses with the emergence of subgap states (Fig. 3b), resulting in local stripe modulations in the gap map (Fig. 3c). Cross-correlation analysis suggested that the subgap states are mainly induced by the twofold defects (Extended Data Fig. 6). By checking the atomic positions of these defects (inset of Fig. 3d), we attribute them to V-site defects (with a concentration of 0.5% yielded by counting), most likely V vacancies, which are two-layer (K and Se) underneath. The local crystallographic environment of a V site naturally leads to anisotropy along either the $x$ or $y$ direction, and we henceforth term these two sublattices $V_x$ (blue) and $V_y$ (red), respectively.

Close examination of high-quality topographies reveals an intriguing fact: these defects are in fact arrow shaped and further classified into four types (Fig. 3d): arrows pointing left (right) along $x$, labelled $V_{xa}$ ($V_{xb}$), and those pointing up (down) along $y$, labelled $V_{ya}$ ($V_{yb}$). Thus, there are two inequivalent sites for each $V_x$ or $V_y$ sublattice, indicating a new periodicity beyond the pristine lattice. These atomically registered defects allow us to establish the spatial correlation among them, which reveals that the two inequivalent sites of $V_x$ or $V_y$ are alternately distributed along the diagonal direction (marked by pink and orange lines in Fig. 3e and f). This explicit arrangement doubles the original unit cell to a new $\sqrt{2} \times \sqrt{2}$ supercell (Fig. 3g). We stress that this supercell represents an intrinsic modulation of the V lattice but is not a simple consequence of the surface K reconstruction because the spatial correlation is long ranged regardless of the surface domains and appears even on the $2 \times 1$ reconstructed surfaces (Extended Data Fig. 5). We attribute this $\sqrt{2} \times \sqrt{2}$ modulation to a density wave order (40,64-66), whose energy gap is strongly suppressed above 100 K (Extended Data Fig. 5). Its spatial pattern (Fig. 3g) potentially breaks the original local rotational symmetry (consistent with the arrow shape of defects) but retains one mirror symmetry of $M_{1\bar{1}0}$, thus revealing a modulation for a 2D $d$-wave altermagnetic order.

**Spin-polarized quasiparticle excitations for opposite-magnetic sublattices**

V-site defects offer an opportunity to study the underlying electronic structure. They generate unidirectional quasiparticle interference (QPI) standing waves that extend several nanometers along $x$ or $y$ in the $dI/dV$ maps (Fig. 4a), depending on which sublattice they are located. The Fourier transformations (FTs) reveal two main sets of scattering vectors, $\boldsymbol{q}_x^1$ and $\boldsymbol{q}_y^1$, each of which consists of multiple parallel stripes in $\boldsymbol{q}$-space that reflect the morphology and dispersion of the constant-energy contour (CEC) in $\boldsymbol{k}$-space (Fig. 4b). A comparison with the calculated CEC near $E_F$ suggests that $\boldsymbol{q}_x^1$ ($\boldsymbol{q}_y^1$) connects the quasi-1D band along the $\boldsymbol{q}_x$($\boldsymbol{q}_y$) direction (Fig. 4c), and the extracted energy dispersion (Fig. 4d) well matches the theoretical calculations with in-plane Néel AFM order (Extended Data Fig. 7). Additionally, the $\sqrt{2} \times \sqrt{2}$ density wave ($\boldsymbol{Q}^*$) folds the bands and opens an energy gap of approximately 20 meV in the energy spectrum of $\boldsymbol{k}$-space eigenstates, with a pronounced spectral weight occurring at the edges of this gap (Fig. 4d).

These defects also give rise to subgap quasiparticle states that spatially oscillate with period $\lambda \cong \pi/k_F$ in their vicinity (Fig. 4e and f). We make use of these states to uncover the spin polarization of these defects. To this end, we choose spatially isolated defects to minimize the influence of other nearby defects and apply magnetic fields perpendicular to the sample. The energy difference for these subgap states is approximately 0.9 meV for magnetic fields of $\pm 7$ Tesla (Fig. 4g and h), yielding a local moment of $1.13\mu_B$, a value close to that of a V atom extracted from neutron diffraction (67). Notably, these peaks shift in opposite directions with magnetic fields for defects on $V_x$ and $V_y$, indicating opposite spin polarizations of these two sublattices (more data in Supplementary Note 5). This finding reveals an in-plane AFM order, and provides direct evidence for unidirectional, SP quasiparticle excitations around two different sublattices.

**Visualization of *d*-wave altermagnetic order in momentum space**

Next, we turn to the spin-related characteristic in ***k***-space. For spin-summed tunneling, the scattering events (Fig. 4a) and low-energy excitations (Fig. 4f) on the $V_x$ and $V_y$ sublattices (with opposite spins) are equally imaged (Fig. 5a). Therefore, in a large field of view (FOV) with randomly distributed defects and using an isotropic PtIr tip (Supplementary Note 4), the QPI intensities at $\boldsymbol{q}_x^1$ and $\boldsymbol{q}_y^1$ show negligible differences (Fig. 5b, c). However, the situation is different when a helical-edge-mode tip is used. At opposite biases, the 'helical tunneling' of the edge modes selects quasiparticles with opposite spins, resulting in highly anisotropic signatures between the ***x*** and ***y*** directions (Fig. 5d). More importantly, this anisotropy lies in the SP electronic modulations for $V_x$ and $V_y$ (Fig. 5e) and will rotate $\pi/2$ when the bias polarity reverses, which can serve as a "smoking gun" proof for *d*-wave altermagnetic spin splitting.

We perform SP-QPI measurements in FOVs of typical 50-nm square, and present three sets of resulting maps in Fig. 5 f, h and Extended Data Figs. 8 to 10. A visual inspection of the FTs reveals a marked intensity anisotropy for $\boldsymbol{q}_x^1$ and $\boldsymbol{q}_y^1$. For example, $\boldsymbol{q}_x^1$ appears to be more intense than $\boldsymbol{q}_y^1$ at positive biases (Fig. 5f and Extended Data Fig. 8). The anisotropy can be quantified by comparing FT linecuts along $\boldsymbol{q}_x$ and $\boldsymbol{q}_y$: while the intensity of $\boldsymbol{q}_y^1$ is weak and even hard to distinguish from the background, the signal of $\boldsymbol{q}_x^1$ is obvious (Fig. 5g). The key observation is that this intensity anisotropy does rotate $\pi/2$ at negative biases (Fig. 5h, i), with the dominant signal now appearing at $\boldsymbol{q}_y^1$. To clarify the ***r***-space signatures, we suppress the long-range randomness in the *dI/dV* maps by Fourier filtering the signals at $\boldsymbol{q}_x^1$ and $\boldsymbol{q}_y^1$ equally (by choosing identical windows for $\boldsymbol{q}_x^1$ and $\boldsymbol{q}_y^1$, as shown in Fig. 5f, h). Standing waves dominant along ***x*** can be directly observed at positive bias, whereas those dominant along ***y*** are observed at negative bias (inset of Fig. 5g, i and Extended Data Figs. 9c, d), confirming that helical tunneling indeed selects quasiparticles on $V_x$ and $V_y$ with opposite spins. We repeated the experiment on multiple samples with different tips (Methods) and were able to find domains in which $V_x$ and $V_y$ are relatively rotated by $\pi/2$ in one sample with the sample tip (Extended Data Fig. 10).

We then create a 2D magnetic contrast map for each QPI image $I(\boldsymbol{q}, eV)$ by defining $M(\boldsymbol{q}, eV) = (I(\boldsymbol{q}, eV) - R_{\pi/2} I(\boldsymbol{q}, eV))/I(\boldsymbol{Q}_{Bragg}, eV)$, where $R_{\pi/2}$ is the rotation operation of $\pi/2$ and $I(\boldsymbol{Q}_{Bragg}, eV)$ is a normalization factor employing the intensity of surface Bragg peaks that reflects the atomic modulation. The opposite signs along the ***x*** and ***y*** axes in the $M(\boldsymbol{q}, eV)$ maps are clearly observed and reversed at opposite biases (Fig. 5j and k), which is consistent with theoretical simulations (Fig. 5e) and serves as direct evidence for *d*-wave altermagnetic spin polarization in momentum space.

We note that the simultaneous detection of opposite-spin states at positive and negative biases in a single measurement provides a robust approach for determining the alternating spin-splitting between the X and Y valleys. Crucially, the observed bias-dependent selection of quasiparticles and the rotation of QPI patterns cannot be attributed to artifacts from an anisotropic tip apex or tip changes; this decisively rules out the significant complications that usually hinder a solid conclusion when using traditional magnetic tips.

**Discussion and Outlook**

By simultaneously measuring spin-resolved dual-space features with a single probe, our results unambiguously demonstrate a 2D *d*-wave altermagnetic order in $KV_2Se_2O$. For bulk $KV_2Se_2O$, theoretical calculations suggest either a C-type (ferromagnetic along the *z* axis) or a G-type (antiferromagnetic along the *z* axis) AFM ground state with nearly degenerate energies due to weak interlayer coupling (Supplementary Note 7). Since the electronic structure of this layered compound is highly two-dimensional (40, 41), a C-type AFM order would lead to a bulk *d*-wave altermagnet, whereas the G-type case can be regarded as vertical stacking of 2D *d*-wave altermagnetic layers with alternating in-plane spin polarizations. We note that we have not been able to locate single unit-cell step edges on potassium-ordered surfaces, possibly because the easy exfoliation of this layered crystal results in large, flat terraces, leaving stacking order along the *c* axis to be an open question for future studies. While recent neutron diffraction data favors the G-type order (67), it is possible that a change in the K concentration (63), or a density-wave modulation along the *c* axis (67), might alter the magnetic ground state that allows for a bulk *d*-wave altermagnetic order. In any case, the discovery of 2D *d*-wave altermagnetic order in a layered material holds great promise for exploring new physics and developing few-layer spintronic devices based on altermagnetism (29-33).

Moreover, our proof-of-principle experiments represent a direct implication of topological materials in modern techniques. Compared to traditional magnetic STM tips, this new approach has profound practical advantages. First, as the helical edge mode is topologically protected, the inevitable tip-sample interactions (such as gentle scratches) usually do not affect spin polarization. In addition, the tips do not have net magnetization and hence are insensitive to the external magnetic field. These two advantages enable high-reliable SP-QPI imaging in large FOVs for *k*-space information (Fig. 5) irrespective of the external magnetic field, which significantly broadens the application scope of spin-STM. Second, it does not require complicated processes for training and preserving the magnetization at the tip apex. As the helical edge modes in $Bi_4Br_4$ are nearly $S_z$ quantized (60,61), the direction of spin polarization can, in principle, be determined by the orientation of the tips' crystal structure. The effective polarization ratio, defined as $P = (I(\boldsymbol{q}_x^1) - I(\boldsymbol{q}_y^1))/(I(\boldsymbol{q}_x^1) + I(\boldsymbol{q}_y^1))$, reaches approximately $\pm 12\%$ in our experiments (Fig. 5l). This value varies among different tips, possibly due to the specific geometry of the tip apex, because tunneling involving two branches of helical edge modes would cancel each other and lead to reduced net spin polarization (Fig. 1c). As a result, the polarization ratio can be further improved by fabricating asymmetric $Bi_4Br_4$ tips in a controlled manner. Finally, the large bulk insulating gap of the $Bi_4Br_4$ tips allows for the investigation of magnetic phenomena in a large energy window and at elevated temperatures. We believe that this technique would be a primary method for probing the atomic-scale spin textures of complicated 2D magnets and moiré materials.

Building on the insights of our work, we envision that our success in nanofabricating tips from bulk quantum material will open up an exciting direction in the functionalization of STM tips to directly visualize quantum phenomena at the atomic scale.

# Figure 1

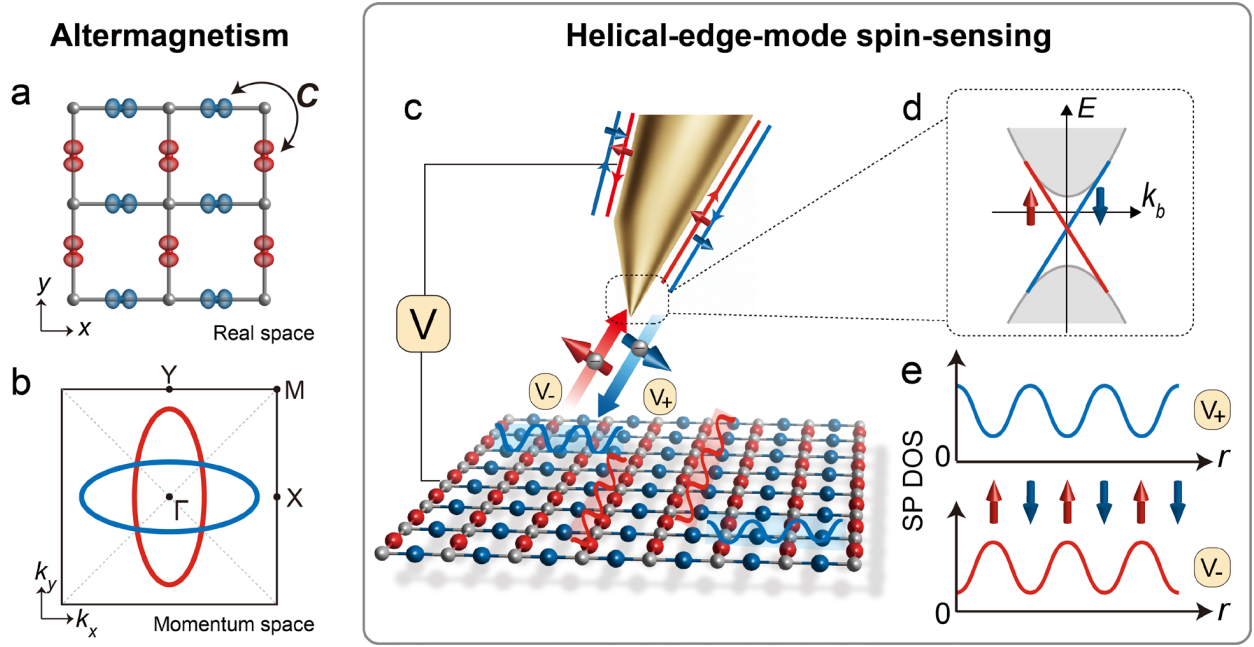

**Figure 1 Schematics of d-wave altermagnetism and the concept of helical-edge-mode spin sensing. a-b**, Real-space lattice and alternating spin densities (**a**) and ***k***-space SML Fermi surface (**b**) of a *d*-wave altermagnet. Here the two spin-opposite sublattices are linked by crystal symmetries (***C***, mirror or rotation), which enforces opposite spin polarization at crystal-symmetry-related momentum points ***k*** and ***Ck***, leading to ***C***-paired spin–momentum locking. The red and blue colors denote spin-up, and spin-down states, respectively. **c-e**, Schematics of STM tunneling with a helical-edge-mode tip. For a nonmagnetic helical-edge-mode tip, $\mathcal{T}$-paired SML (**d**) ensures that electrons flowing in opposite directions carry opposite spins, resulting in tunneling-direction-dependent spin polarization, which can be controlled by applying a negative or positive bias to the sample (**e**).

# Figure 2

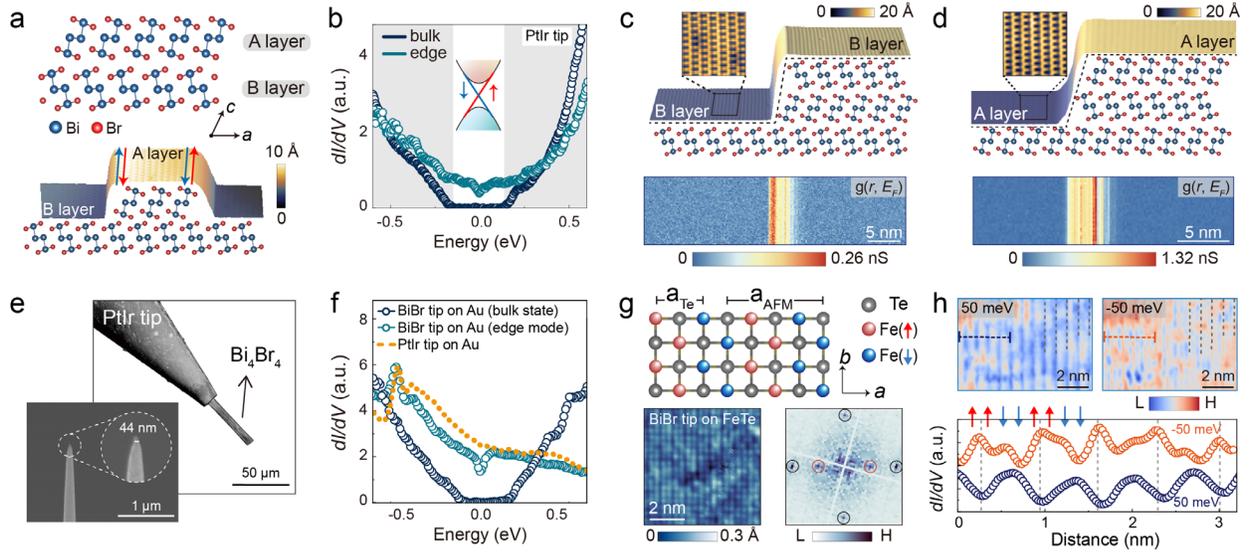

**Figure 2 Characteristics of bulk Bi$_4$Br$_4$ and the helical-edge-mode tips. a,** Crystal structure of α-Bi$_4$Br$_4$ (top) and STM topography of an AB-type monolayer step (bottom) with helical edge modes located at the step edge. **b,** Typical $dI/dV$ spectra of the bulk state (dark blue) and the edge states (light green) in the α-Bi$_4$Br$_4$ crystal measured with PtIr tips. **c-d,** Topography (top) and $dI/dV$ map at E$_F$ (bottom) for a BB-type bilayer step (**c**) and an AA-type bilayer step (**d**), showing the edge modes in both cases. **e,** High-resolution SEM image of a Bi$_4$Br$_4$ (BiBr) tip bonded to a PtIr wire. The inset shows the tip apex at higher magnification. **f,** $dI/dV$ spectra obtained on the Au (111) surface using a BiBr tip. The dark blue curve corresponds to the bulk-state tip, and the light green curve corresponds to the helical-edge-mode tip. Both curves resemble those shown in **b**, except for the peak at -0.5eV due to the surface state of Au (111) (orange dots). **g,** Atomic and spin structures of Fe$_{1.02}$Te (top), and the topography (bottom left) and its FT (bottom right) measured with a helical-edge-mode tip. Bragg peaks and AFM vectors are marked by black and red circles, respectively. **h,** $dI/dV$ maps at ±50 mV in the same FOV. Line profiles along the dashed lines show a contrast reversal at opposite biases, indicating that the reversed direction of tunneling detects the opposite-spin-polarized density of the sample. STM setup conditions: $V_s$ = -0.6 V, $I_t$ =0.1 nA (**a**); $V_s$ = -0.7 V, $I_t$ = 1 nA, bias modulation $V_{mod}$ = 5 mV (**b**); $V_s$ = -0.6 V, $I_t$ = 1 nA, $V_{mod}$ = 10 mV (**c-d**); $V_s$ = 1 V, $I_t$ = 1 nA, $V_{mod}$ = 10 mV (**f**); $V_s$ = 15 mV, $I_t$ = 0.7 nA (**g**); $V_s$ = 50 mV, $I_t$ = 0.8 nA, $V_{mod}$ = 5 mV (**h**).

# Figure 3

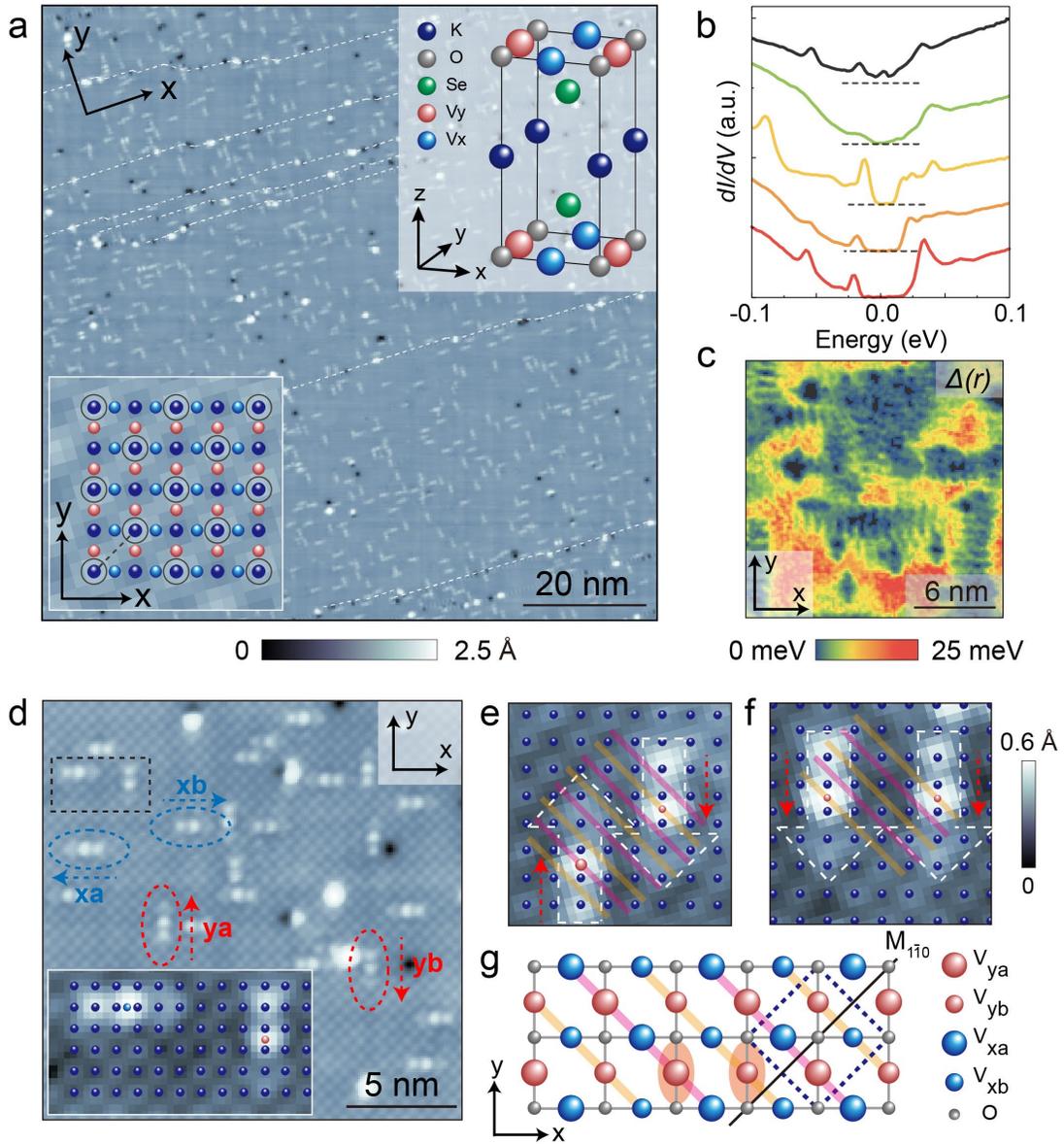

**Figure 3 Crystal symmetry and density wave state in KV$_2$Se$_2$O. a,** Typical topographic image of KV$_2$Se$_2$O with $\sqrt{2} \times \sqrt{2}$ reconstruction of K atoms. Top-right inset: crystal structure; bottom-left inset: topography with a transparent overlay showing the relationship between the locations of K, O and V atoms. The white dashed lines mark domain walls for surface reconstruction (Extended Data Fig.5). **b,** Representative $dI/dV$ spectra showing spatial variations. **c,** Spatial-resolved gap map on a 19-nm square FOV. **d,** Topography showing four different types of arrow-shaped defects (inset: an enlarged view of the region marked by the dashed rectangle, showing the atomic location of the defects). **e-f,** Spatial correlation among these defects. Two inequivalent sites (for example, $V_{ya}$ and $V_{yb}$) are alternately distributed along the diagonal direction and are marked by pink ($V_{ya}$) and orange ($V_{yb}$) lines, respectively. **g,** The demonstrated real-space pattern resulting in a new $\sqrt{2} \times \sqrt{2}$ supercell (dashed square), with one remaining $M_{1\bar{1}0}$ mirror plane (dark line). STM setup conditions: $V_s = 0.4$ V, $I_t = 20$ pA (**a**); $V_s = 0.4$ V, $I_t = 1.5$ nA, $V_{mod} = 4$ mV (**b-c**); $V_s = 0.4$ V, $I_t = 20$ pA (**d-f**).

# Figure 4

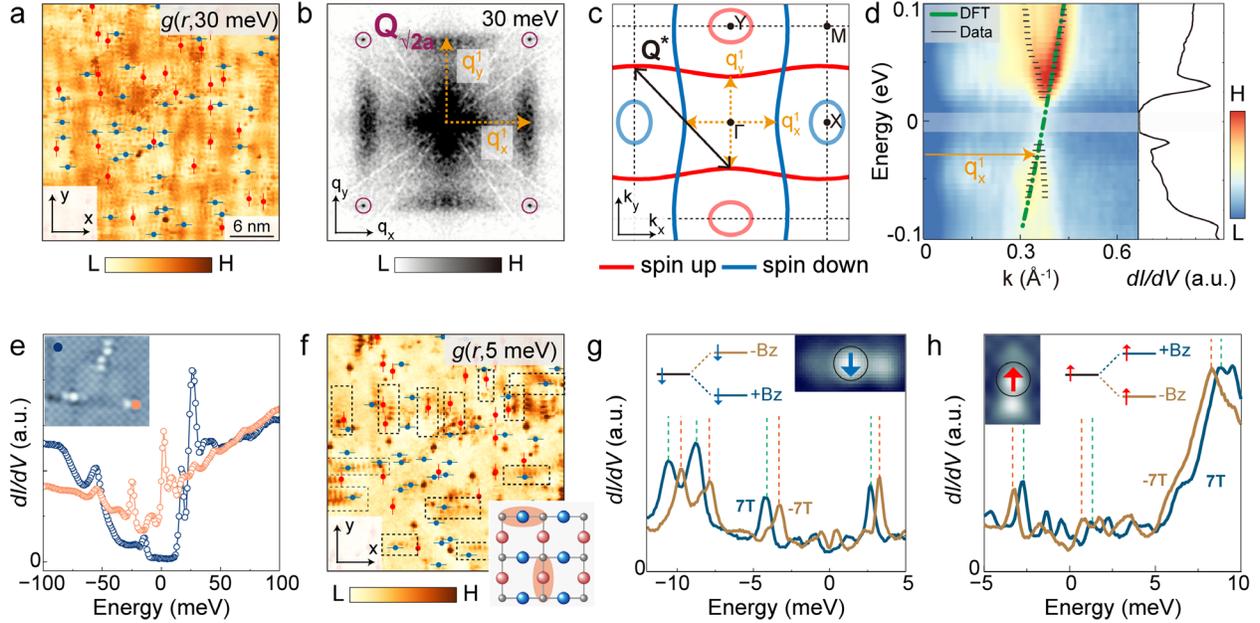

**Figure 4 Quasiparticle interference and spin-polarized subgap excitation. a**, 46 nm ×46 nm $dI/dV$ map at 30 mV. The blue and red dots denote V-site defects on $V_x$ and $V_y$, respectively. **b**, $I(\mathbf{q}, eV)$, the FT of **a**, showing two main quasi-1D scattering signals at $\mathbf{q}_x^1$ and $\mathbf{q}_y^1$. The purple circles denote the surface Bragg peak with $\sqrt{2}\times\sqrt{2}$ reconstruction. **c**, Calculated Fermi surface, with $\mathbf{q}_x^1$ and $\mathbf{q}_y^1$ labeling inter-band scattering. $\mathbf{Q}^*$ denotes the nesting vector of the density wave, which is the same as the surface Bragg vector. **d**, Energy dispersion of the band extracted from the QPI along $\mathbf{q}_x$. The green dashed line denotes the calculation, and the dark dots track the maximal intensity of the experimental data, which clearly shows a band folding and the opening of an energy gap. Note that the back-folding points are not exactly located at $k_F$, a typical feature for density-wave order. **e**, Typical $dI/dV$ spectra of the defect-free region (blue) and near a defect (orange). **f**, A $dI/dV$ map at 5 mV in the same FOV of **a**, showing unidirectional modulations of the subgap states on $V_x$ and $V_y$. **g, h**, Magnetic-field-dependent peak shifts in the $dI/dV$ spectra, for $V_{xa}$ (**g**) and $V_{ya}$ (**h**) defects. The opposite direction of the shifts indicates opposite spin orientations for the $V_x$ and $V_y$ sublattices. STM setup conditions: $V_s$ = 0.4 V, $I_t$ = 1 nA, $V_m$ = 5 mV (**a-b, d, f**); $V_s$ = 0.2 V, $I_t$ = 2 nA, $V_{mod}$ = 2 mV (**e**); $V_s$ = 0.1 V, $I_t$ = 2 nA, $V_{mod}$ = 0.2 mV (**g-h**).

# Figure 5

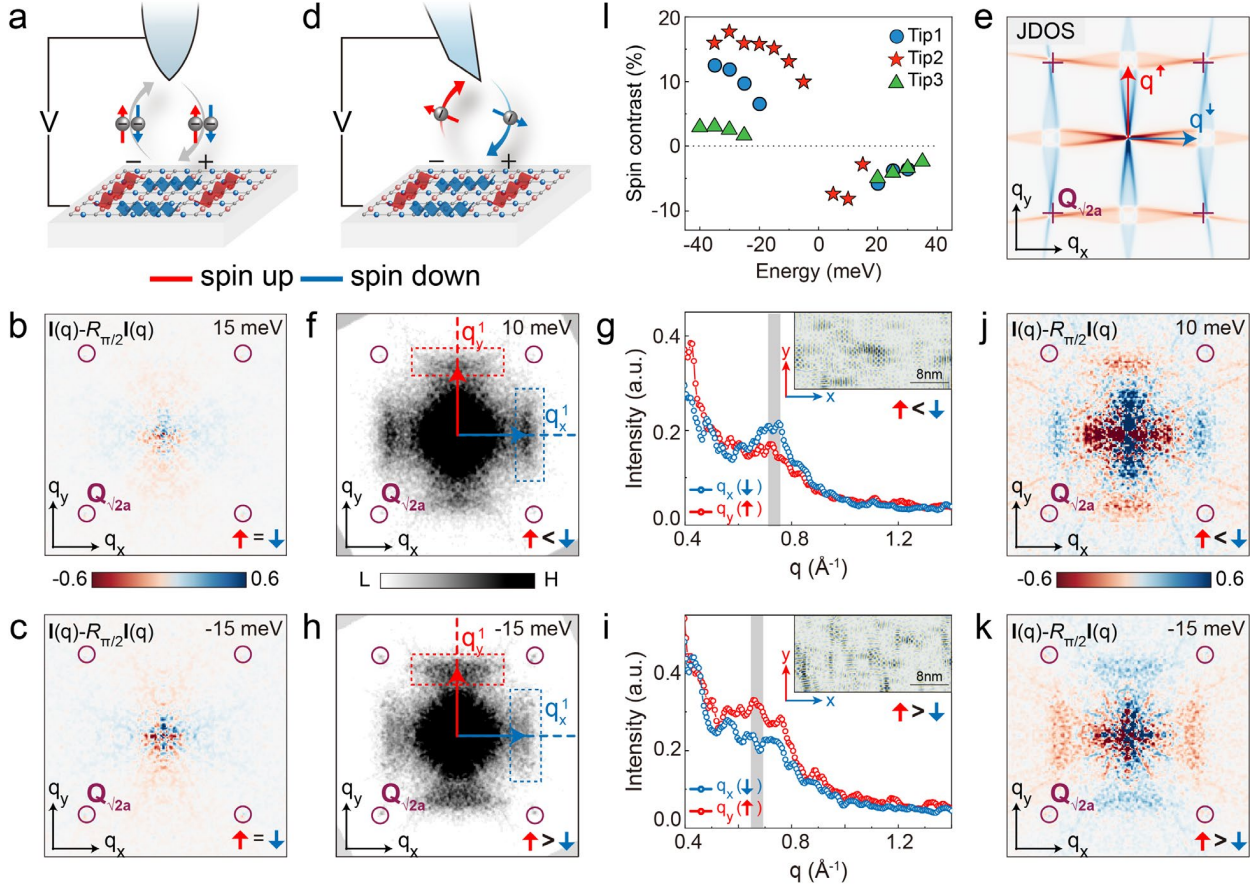

**Figure 5 Visualization of the d-wave altermagnetic order. a,** Schematic of spin-summed tunneling with a normal metallic tip, in which the spin-polarized modulations along *x* and *y* can be equally imaged. **b, c,** Magnetic contrast map $M(\boldsymbol{q}, eV)$ (defined in the main text) at 15 mV (**b**) and -15 mV (**c**), measured by a PtIr tip. **d,** Schematic of helical tunneling in which the flow direction of the tunneling current selects different spin-polarized states. **e,** Theoretical spin-resolved joint density of states (JDOS) at the $E_F$. **f, h,** QPI images at 10 mV (**f**) and at -15 mV (**h**) showing a reversal of the anisotropy in the QPI intensity at $\boldsymbol{q}_x^1$ and $\boldsymbol{q}_y^1$, using a helical-edge-mode tip. This is because the spin-polarized electronic states of $V_x$ and $V_y$ can be mapped by spin-selective tunneling at opposite biases. **g, i,** Corresponding FT linecuts along the $\boldsymbol{q}_x$ (blue) and $\boldsymbol{q}_y$ (red) directions for **f** and **h**, respectively. The inserts show the quasiparticle modulations in real-space by equal Fourier filtering $\boldsymbol{q}_x^1$ and $\boldsymbol{q}_y^1$ in the same relevant *q*-space windows marked by dashed rectangles in **f** and **h**. However, the dominant modulations appear along *x* (*y*) at positive (negative) bias. **j, k,** Spin-resolved magnetic contrast map $M(\boldsymbol{q}, eV)$ extracted from **f** and **h**, respectively. The opposite signs along the *x* and *y* axes serve as direct evidence for *d*-wave spin polarization, and the signs reverse at opposite biases. **l,** Effective polarization ratio (see main text) for three BiBr tips. STM setup conditions: $V_s$ = 0.4 V, $I_t$ = 1 nA, $V_{mod}$ = 5 mV.

# Method

**Topologic classification and helical edge modes in α-Bi₄Br₄**

Helical edge modes are the hallmark of a 2D quantum spin Hall insulator featuring a $\mathbb{Z}_2 = 1$ topological index, whose discovery has triggered major breakthroughs in understanding topological materials over the past decade (53,54). Further developments in topological theory have proposed that helical edge modes can also emerge in 3D bulk crystals, for example, at certain hinges in higher-order (HO) topological insulators (TIs), with the $\mathbb{Z}_4$ indicator in centrosymmetric space groups (55, 56), or at the boundary of each layer in weak TIs (68), with at least one of the three weak indices $z_{2w,w=1,2,3}$ being nonzero. However, the number of materials realizing these topologies is still quite limited.

Bi₄Br₄ is a quasi-1D van der Waals material composed of [Bi-Br]ₙ atomic chains aligned along the crystallographic ***b***-axis (57), and crystalized in a centrosymmetric monoclinic lattice (space group C2/m) with inversion symmetry within the layer. Monolayer Bi₄Br₄ is a 2D QSHI (57) with 1D helical edge modes within the bulk gap (see Extended Dada Fig. 1). For bulk α-Bi₄Br₄, one unit cell comprises two AB stacked layers along the *c*-axis (Fig. 2a); thus, the inversion center is not in the middle of a bilayer but rather in a monolayer. From the perspective of symmetry indicators, it is a HOTI with a topological indicator $\mathbb{Z}_4 = 2$. As a consequence, *gapless* helical edge (hinge) states emerge on certain hinges of the crystals depending on the stacking geometry (58,59), whereas hybridization occurs on the other hinges (see Supplementary Note 2).

On the other hand, new concept of a 3D QSHI has been recently established (60,61) to characterize the intrinsic spin-resolved topology of these 3D insulators. From the viewpoint of spin resolved topology, a 3D QSHI is defined by nonzero spin-resolved Chern numbers (i.e., $C_{sz} = 2$) for all the $k_z$ planes, resulting in a non-quantized (but generally nonvanishing) $S_z$ spin Hall conductivity per bulk unit cell. It has been shown that the helical HOTIs (inversion $\mathbb{Z}_4 = 2$) can host three types of spin-resolved topologies: the $\mathcal{T}$-doubled axion insulator, 3D QSHI and spin Weyl phase; in the $S_z$-resolved topology, the helical HOTI α-Bi₄Br₄ is classified to the 3D QSHI phase (60). Phenomenally, a 3D QSHI is similar to a weak TI: they have accumulating helical edge states on the side surfaces, and one 2D layer contributes one helical edge mode. When the interlayer interaction is weak and the spin $S_z$ is nearly conserved, the spin Hall conductivity is expected to be nearly quantized, with the edge states being slightly *gapped* in the 3D QSHI. As shown in Fig. 2 and Extended Dada Fig. 1, helical edge modes are observed at different types of bilayers (despite their different intensities). This finding is indeed consistent with the prediction for a 3D QSHI (60,61). Therefore, α-Bi₄Br₄ can be catalogued as a 3D QSHI ($C_{sz}^{\pm} = \pm 2$), with *gapless* helical edge states on certain hinges and *slight-gapped* helical edge states on the rest, and it could host a nearly quantized spin Hall conductivity of 3.62×e/4π (60) per layer (see Supplementary Note 2 for more details). In principle, both of these types of edge modes can contribute to the quantum tunneling of electrons in our experiments.

**Fabrication and calibration of the Bi₄Br₄ tip**

Single crystals of Bi$_4$Br$_4$ were synthesized through a chemical vapour transport reaction. The tips were fabricated in a focused ion beam-scanning electron microscope (FIB–SEM) dual-beam system. We present the procedure in Extended Dada Fig. 2. First, a tiny bulk Bi$_4$Br$_4$ sample with a thickness of typically several micrometres was selected and secured on carbon tape. A slender strip (approximately 50 μm × 5 μm × 5 μm) was then patterned via gallium ion-beam etching, with one end affixed to the pristine crystal. The other end of this strip was then affixed to an omniprobe by depositing a Pt bonding pad, after which the sample was detached from the crystal by gently retracting the probe while etching the joint end. Second, the Bi$_4$Br$_4$ strip was transferred and bonded to a prefabricated PtIr tip. To ensure stable attachment, the tip apex of PtIr was milled to be flat. The Bi$_4$Br$_4$ strip was aligned appropriately and placed at the center of the top platform of the PtIr tip, and Pt bonding pads were deposited at the interface to securely bond the strip on the top of the PtIr tip. Finally, the tip was sharpened by gentle ion-beam milling to obtain a sharp apex with a curvature diameter of less than 100 nm (Extended Data Fig. 2).

The obtained Bi$_4$Br$_4$ tip was always first calibrated on the Au (111) surface at 4. 8K. Two types of spectra, referred to as the bulk-state tip and helical-edge-mode tip, are obtained, as shown in Fig. 2f and Extended Data Fig. 3. By comparing the spectra of the PtIr-tip@Bi$_4$Br$_4$-crystals and Bi$_4$Br$_4$-tip@Au(111), we conclude that the gentle gallium ion-beam etching during the fabrication process does not significantly alter the properties of Bi$_4$Br$_4$. First, etching does not induce doping because the bulk gap of a similar size still scans the Fermi level on the Bi$_4$Br$_4$ tip. Second, the featured suppression of the local density of states (DOS) at E$_F$ for the helical edge modes also does not change. Specifically, this zero-bias anomaly is an indicator of Tomonaga-Luttinger liquid (TLL), a typical electronic-correlated state in 1D. We fit this low-energy spectral feature with the TLL model of $\rho_{TLL}(\epsilon, T) \propto T^\alpha \cosh\left(\frac{\epsilon}{2k_BT}\right)$ (69, 70) and find very similar behavior for the 1D edge modes in the bulk crystal Bi$_4$Br$_4$ and Bi$_4$Br$_4$ tips (Extended Data Fig. 3b-e). These observations not only suggest that the electronic structures of Bi$_4$Br$_4$ remain unchanged on the tips but also prove that the tunneling is indeed mediated by helical edge modes. The peak at -480 mV is related to the surface state on the Au (111) surface.

The two statuses of a tip, that is, the bulk-state and helical-edge-mode types, can be switched *in situ* by applying large electronic pulses (Extended Data Fig. 3e, f). We achieve helical-edge-mode tips in this way on the Au (111) surface and check its TLL feature near E$_F$. For these helical-edge-mode tips, excellent atomic resolution can be achieved (Extended Data Fig. 3g-l).

In this work, we made more than twenty Bi$_4$Br$_4$ tips. Approximately half of them can be properly treated and calibrated on the Au (111) surface with decent spatial resolution and accurate spectroscopic features (others frequently crash to the sample). Two of these tips were used to image FeTe, and both tips showed atomic resolution and spin resolution (Supplementary Note 3). Because KV$_2$Se$_2$O usually has uneven, complex and disordered surfaces, performing STM experiments is more difficult. Ten of them were used to study KV$_2$Se$_2$O single crystals. Most of these tips can achieve atomic resolution if they land on ordered surfaces, but only five of them show clear spin resolution. This may be because the spin orientation for the edge modes is 'horizontal' (Fig. 1c), whereas that for the AFM order in KV$_2$Se$_2$O is out-of-plane.

**Single-crystal growth and characterization of KV$_2$Se$_2$O**

KV$_2$Se$_2$O single crystals were obtained through a three-step synthesis. First, the precursor VSe$_2$ was synthesized by annealing stoichiometric amounts of V (99.9%, powder) and Se (99.9%, powder) at 700 °C for 24 h. Polycrystalline samples of KV$_2$Se$_2$O were then synthesized *via* a solid-state reaction from a mixture of K (99%, lump), V$_2$O$_5$ (99.9%, powder), V (99.9%, powder), and the pre-synthesized VSe$_2$, with a molar ratio of K : V : Se : O = 1.1 : 2 : 2 : 1. The mixture was sealed in an alumina crucible inside a quartz ampoule, which was evacuated to high vacuum using a mechanical pump before sealing, heated to 1000 °C, held for 15 h, and finally furnace-cooled. The as-synthesized polycrystalline samples were used as the starting material for single-crystal growth of KV$_2$Se$_2$O via the self-flux method, employing KSe as the flux. The reactants were loaded into an alumina crucible, which was then sealed in a niobium tube and subsequently encapsulated in an evacuated quartz ampoule. The quartz ampoule was heated to 1000 °C, held at this temperature for 20 h, slowly cooled to 650 °C over 7 days to optimize the crystal quality, and finally allowed to cool naturally to room temperature. Upon opening the quartz ampoule and niobium tube, black single crystals of high quality were obtained. These crystals exhibited pronounced sensitivity to air and moisture, requiring all handling operations to be performed in an inert-atmosphere glove box.

The single-crystal structure was examined via X-ray diffraction (XRD) using a Rigaku SmartLab 9 X-ray diffractometer with Cu $K_\alpha$ radiation. The chemical composition of the single crystal was characterized via energy-dispersive X-ray spectroscopy (EDS) using a ZEISS EVO-MA10 scanning electron microscope equipped with an Oxford Instruments X-act spectrometer. The average K : V : Se atomic ratio is displayed in Extended Data Fig. 4. In our case, the concentration of V vacancies (defects) was determined to be 0.5% by counting the atomic defects in multiple FOVs. This information together with the EDS data, suggests that the K:V:Se atomic ratio is 0.975:1.995:2.023, which is very close to 1:2:2.

The transport properties were measured via a Quantum Design Physical Property Measurement System (PPMS). Owing to the sensitivity of KV$_2$Se$_2$O to air and moisture, the crystals were cleaved in an argon-filled glovebox. Electrodes were fabricated on a freshly cleaved surface using conductive silver paste and platinum wires, and mounted on a sealed sample puck to prevent degradation during the measurements. The temperature-dependent in-plane resistivity $\rho$(T) shows weak insulating behavior at low temperatures (Extended Data Fig. 4), in contrast to a previous report (64), but is consistent with our STM measurements that revealed an energy gap at E$_F$ (Fig. 4d).

**STM experiments**

The majority of the STM data were acquired with a commercial CreaTec low-temperature STM at 4.8K, and the sub-kelvin studies were performed on a Unisoku USM1300 system. The KV$_2$Se$_2$O single crystals were cleaved *in situ* in cryogenic ultrahigh vacuum at $T \approx 30$ K and immediately inserted into the STM head. All the PtIr and Bi$_4$Br$_4$ tips were calibrated on a single-crystal Au (111) surface prior to the measurements. Differential conductance (*dI/dV*) spectra were acquired via the standard lock-in technique at a frequency of 987.5 Hz under a modulation voltage of 0.2–5 mV. The data measured with PtIr tips are shown in Supplementary Note 4.

The low-energy local density of states (LDOS) in KV$_2$Se$_2$O exhibits significant spatial inhomogeneity. To map the spatial distribution of the gap size, we performed *I–V* spectroscopy with a 136 × 136-pixel grid on a 19-nm square FOV (Fig. 3c). For each spectrum, the gap size was determined by identifying the energy range where the tunneling current falls below the near-zero threshold (defined as current < 0.3 pA). The extracted gap values were then compiled to generate a nanoscale-resolved gap map. For QPI imaging, we typically measured *dI/dV(r, eV)* maps in 50-nm square. The Lawler-Fujita drift-correction algorithm was used (71). We twofold symmetrized the obtained FTs along $q_x$ or $q_y$, and the resulting maps were rotated to align the lattice direction in the same way for each dataset and then cropped to highlight the main features in the first Brillouin zone.

To better illustrate the anisotropy in the QPI intensity along the $q_x$ and $q_y$ directions, we compared the original twofold symmetrized QPI pattern with its 90° clockwise-rotated counterpart. This comparison is performed by introducing a spin-resolved magnetic contrast map for each QPI image, defined as $M(q, eV) = (I(q, eV) - R_{\pi/2}I(q, eV))/I(Q_{Bragg}, eV)$, as shown in the main text (Supplementary Note 6). For the dataset measured with PtIr tips, $M(q, eV)$ exhibited negligible differences in the intensities at $q_x^1$ and $q_y^1$, which is consistent with spin-summed tunneling (Fig. 5a-c and Supplementary Note 4). In contrast, the data with BiBr tips reveal a pronounced anisotropy that reverses at opposite biases (Fig. 5f-k). The effective spin polarization ratio is defined as $P = (I(q_x^1) - I(q_y^1))/(I(q_x^1) + I(q_y^1))$, where $I(q_x^1)$ and $I(q_y^1)$ denote the average QPI intensities at $q_x^1$ and $q_y^1$, respectively. To extract $I(q_x^1)$ and $I(q_y^1)$, we choose two small, relevant *q*-space windows for $q_x^1$ and $q_y^1$ with identical window sizes, and integrate the signals within each window. Owing to the energy-dependent dispersion of the QPI features, the positions of these analysis windows were adjusted accordingly at different energies.

**DFT calculations**

The calculations were performed in the framework of density functional theory as implemented in the Vienna ab initio simulation package (72). The projector-augmented wave potential was adopted with the plane-wave energy cut-off set at 440 eV. The exchange-correlation functional of the Perdew–Burke–Ernzerhof type has been used for both structural relaxation and self-consistent electronic calculations, with convergence criteria as $10^{-5}$ eV(72, 73). The GGA+U method was employed to treat the strong correlations of the V d orbitals, where the value of Hubbard U was taken as 1 eV, which provided the best fit for the experimental data. The Brillouin zone was sampled by an 11 ×11 × 5 Γ -centered Monkhorst–Pack mesh. We construct a Wannier tight-binding model Hamiltonian by the WANNIER90 interface (74), with V d orbitals and Se p orbitals, and the DOS and QPI are calculated via the Wannier model with a "201×201×1" k-mesh. More calculation results are shown in Supplementary Notes 7 and 8.

## Acknowledgements

We thank Vidya Madhavan, Quansheng Wu, Lin Jiao, Shichao Yan, Kun Jiang, Tong Zhou and Zhenyu Zhang for valuable discussions. This work is supported by the National Key R&D Program of the MOST of China (Grants No. 2022YFA1602600 and 2021YFA1401500), the Scientific Research Innovation Capability Support Project for Young Faculty (No. ZYGXQNJSKYCXNLZCXM-M25), the Hong Kong Research Grants Council (C6046-24G, 16306722, 16304523 and 16311125), the Quantum Science and Technology-National Science and Technology Major Project (Grant No. 2021ZD0302800), he National Natural Science Foundation of China (Grant Nos. 12488201, 52261135638, 52373309), and the HFNL Self-Deployed Project (ZB2025020200, ZB2025020100). This research was partially carried out at the Instruments Center for Physical Sciences, the USTC Center for Micro and Nanoscale Research and Fabrication.

## Author contributions

Zhenyu Wang, J. L. and X.H.C. conceived the idea and supervised the project. Zhuying Wang, S. Y., W. M., H. S., K. Z., Y. Z., and Y. M. fabricated the tips and performed the STM experiments and data analysis under the guidance of Zhenyu Wang. X.K.C. and J. L. performed the band calculations and theoretical analysis. X. X., X. S. and B. X. synthesized the $KV_2Se_2O$ single crystals. F. Q. characterized the $KV_2Se_2O$ samples. W. L. and P. Y. synthesized and characterized the Bi$_4$Br$_4$ single crystals. Zhenyu Wang and J. L. interpreted the results with assistance from Y. G. and Q. N. Zhenyu Wang, J. L. and X.H.C. wrote the manuscript with input from all the authors.

## Competing financial interests

The authors declare no competing interests.

## Data availability

Source data will be provided with this paper. More data supporting the findings of this study are available from the corresponding author upon reasonable request.

## Code availability

The code used for STM data analysis is available from the corresponding author upon reasonable request.